\documentclass[12pt]{article}
\usepackage[T1]{fontenc}
\usepackage[latin1]{inputenc}
\usepackage{graphics}
\usepackage{latexsym}
\usepackage[english]{babel}
\makeatletter

 \newcommand{\lyxaddress}[1]{
   \par {\raggedright #1 
   \vspace{1.4em}
   \noindent\par}
 }

\makeatother

\begin{document}

\title{Effects of the magnetic fields on the helium white dwarfs structure. }
\author{R. Ma\'{n}ka, M. Zastawny-Kubica, A. Brzezina, I. Bednarek}
\maketitle

\lyxaddress{Department of Astrophysics and Cosmology,
Institute of Physics, University of Silesia, Uniwersytecka 4,
40-007 Katowice, Poland.}

\begin{abstract}

In this paper the effect of the magnetic field on the form of the
equation of state and helium white dwarfs structure are discussed.
The influence of the temperature and magnetic field on white
dwarfs parameters have been investigated. The mass-radius
relations for different parameters were obtained. The occurrence
of unstable branches in the mass-radius relation are presented for
the temperature equals $T=3$ $10^{8}$ $K$ and for different values
of the strength of magnetic field. Theoretical model of the star
with two Landau levels is obtained.
\end{abstract}
PACS numbers { 64, 65, 97 }
\section{Introduction}
Properties of matter in strong magnetic field has been the subject
of investigations in astrophysics of white dwarfs and neutron
stars. It is motivated by the fact that magnetic fields of the
order of $10^{8}$ $Tesla$ - $10^{9}$ $Tesla$ are known to exist in
many cases of white dwarfs and neutron stars. White dwarfs are
compact objects with masses comparable to that of the Sun, radii
of the order of several thousands of kilometers and mean densities
about $10^{9} $ $kg$ $m^{-3} $. These stars no longer burn nuclear
fuel. They are slowly cooling as they radiate away their residual
thermal energy. White dwarfs radii decrease with increasing mass.
In accordance with theoretical
predictions there is a critical value of white dwarf mass known as $%
Chadrasekhar $ $mass $ which is predicted to be
\begin{eqnarray*}
M_{Ch}=5.8 \ Y_{e}^{2} \ M_{\odot}
\end{eqnarray*}
where $Y_{e}$ is the fraction of electrons. This is the pressure
of degenerate electrons which supported the star against collapse.
The electrons are supposed to be degenerate with arbitrary degree
of relativity (in this paper the unit in which $c=\hbar=1$ was
used) $x=k_{F}/m_{e} $ and it various with the density. Helium
ions provide the mass of the star and their contribution to the
pressure is negligible \cite{kot5,mil,glen}.\newline The ions form
a regular lattice which minimizes the energy. Detailed models of
white dwarfs were given by Salpeter (1967) who derived conditions
for their solidification and determined properties of the lattice.
\newline A number of white dwarfs with strong magnetic fields was
discovered (Kemp et al. 1970; Putney 1995; Reimers et al. 1996)
\cite{mil9,mil12,mil10} and extensively studied (Jordan 1992;
Chanmugan 1992)\cite{mil7,mil8}.\newline In about 3-4 $\%$ of all
white dwarfs magnetic fields have been detected ranging from $2$
$10^{-1}$ $Tesla$ up to $10^{5}$ $Tesla$. The magnetic white
dwarfs where a surface magnetic field on the order of $10$ $Tesla$
$\sim$ $10^5$ $Tesla$ and an interior field of $10^5$ $Tesla$
$\sim$ $10^{9}$ $Tesla$ are estimated. The magnetic field may
causes considerable effects on the structure of white dwarfs.
Earlier work of Ostriker and Hartwick (1968) and recent
calculations of Such and Mathews (2000) predicted an increase of
white dwarf radii in the presence of internal magnetic fields
\cite{mil13,mil14}. In this paper the influence of the temperature
and magnetic field on white dwarfs parameters have been
investigated. The mass-radius relations for different parameters
were obtained. \newline This paper is organized as
follows.\newline In Sect.1 are presented the general properties of
white dwarfs. In Sect.2 the employed equation of state (EOS) is
obtained in magnetic fields model with different temperatures.
Finally the EOS is used to determine the equilibrium
configurations of white dwarfs. The interesting fact is the
existence of unstable branches on the mass-radius relation for  temperature $%
T=3$ $10^{8}$ $K$ and different values of the strength of magnetic
field, which indicate that  there are several possible
evolutionary tracks. Finally, in Sect.3 the main implications of
the results are discussed.

\section{The white dwarfs in the magnetic fields model}

This paper presents the model of helium white dwarf in which the
main contribution to the pressure comes from ultra-relativistic
electrons. Corrections from finite temperature and magnetic field
are also included.
 Having made the assumption that the ionised
uniform helium plasma forms the interior of such an object one
can say that the white dwarf matter consists of electrically
neutral plasma which comprises charged ions $(^4He)$ and
electrons.\newline The Lagrangian density function in this model
can be represented as the sum
\begin{equation}
{\mathcal{L}}={\mathcal{L}_e}+{\mathcal{L}_G}+{\mathcal{L}_{ion}}+{\mathcal{L%
}_{QED}},  \label{lagra}
\end{equation}
where ${\mathcal{L}_e}$, ${\mathcal{L}_G}$ describe the electron
and gravitational terms, respectively. The ${\mathcal{L}_{ion}}$
determines the ionized helium and ${\mathcal{L}_{QED}}$ is the
Lagrangian density function of the QED theory.\newline The
electron part of the Lagrangian is given by
\[
{\mathcal{L}_e}=i\overline{\psi _e}\gamma ^\mu D_\mu \psi _e-m_e\overline{%
\psi _e}\psi _e,
\]
where $D_\mu $ is the covariant derivative defining as$\,\,D_\mu
=\partial _\mu -ieA_\mu \,$\thinspace and $e$ is the electron
charge. The Dirac equation for electrons obtained from the
Lagrangian function ${\mathcal{L}}$ has the form
\[
(i\gamma ^\mu D_\mu -m_e)\psi _e=0.
\]
 The vector potential is defined as $A_\mu =\{A_0=0 ,\,A_i\}$
 where
\[
A_i=-\frac 12\varepsilon _{ilm}x^lB_0^m.
\]
 The gauge in which uniform magnetic field B lies along the
z-axis was chosen
\[
B_0^m=(0,0,B_z).
\]
 The magnetization is given by the contribution
of the electrons and depends on the density of particles and
antiparticles.
Different from zero magnetization generates internal molecular field $%
B_z=\mu \mathcal{M}_z\neq 0$. The energy-momentum tensor can be
calculated taking the quantum statistical average
\begin{eqnarray*}
\bar{T}_{\mu\nu}=<T_{\mu\nu}>,
\end{eqnarray*}
where
\begin{eqnarray*}
T_{\mu \nu }=2\frac{\partial {{\mathcal{L}}}}{\partial g^{\mu \nu
}}-g_{\mu \nu }{\mathcal{L}}.
\end{eqnarray*}
 In general in the presence of
finite magnetization the pressure is anisotropic. For the
electrically charge particles, different equations of state for
directions paraller and perpendicular to the magnetic field can
be obtained. Thus the energy-momentum tensor has the form
\begin{equation}
<T_{\mu \nu }>=\left(
\begin{array}{cccc}
\varepsilon =\rho & 0 & 0 & 0 \\
0 & P & 0 & 0 \\
0 & 0 & P & 0 \\
0 & 0 & 0 & P+\mu \mathcal{M}_zB_z
\end{array}
\right).
\end{equation}
This anisotropy in the pressure leads to a magnetostriction effect
in the quantum magnetized gas of charged particles. In the
classical case nonzero magnetization produces a flattening effect
in white dwarfs and neutron stars models \cite{chai} similarly
like in the case of rotating stars \cite{konno} . Einstein
equations (in isotropic case) leads to the standard
Tolman-Oppenheimer-Volkov equations \cite{mil4}. The equations
describing masses and radii of white dwarfs are determined by the
proper form of the equation of state. The aim of this paper is to
calculate the equation of state for helium white dwarfs with the
assumptions of finite temperature and in the presence of magnetic
field.\newline The properties of an electron in external magnetic
field have been studied, for example by Landau and Lifshitz in
1938 \cite{mil15}. In this paper the effects of high magnetic
field on the equation of state of a relativistic, degenerate
electron gas is considered. The motion of free electrons in
homogeneous magnetic field of the strength $B$ perpendicular to
the field is confined by the oscillatory force determined by the
field $B$ and is quantized into Landau levels with the energy
$nB$, $n=0,1,\ldots$. In the case of nonrelativistic electrons the
energy spectrum is given by the relation
\begin{eqnarray*}
E_{n,p_{z}}=n\omega_{c}+\frac{p_{z}^{2}}{2m_{e}}
\end{eqnarray*}
where $n=j+\frac 12+s_z$, the cyclotron energy
$\omega_{c}=eB/m_{e}$, $p_{z}$ is the momentum along the magnetic
field and can be treated as continuous. For extremly high magnetic
field the cyclotron energy is comparably with the electron rest
mass energy and this is the case when electrons became
relativistic. Introducing the idea of the critical magnetic field
strength $B_{c}=m_{e}^{2}/\mid e \mid$ which equals $4.414\times
10^9$ $Tesla$, it is easy to distinguish between nonrelativistic
and relativistic cases, for $B\geq B_{c}$ the relativistic Dirac
equation for electrons have to be used. The obtained dispersion
relation now takes the form
\begin{eqnarray}
E_{n,p_{z}}=\sqrt{p_{z}^2+m_e^2+2eB_zn}.\label{eq1}
\end{eqnarray}
Along the field the motion is free, quasi one-dimensional with the
modified density of states. The electron density of states in the
absence of the magnetic field is replaced by the sum
\[
2\int \frac{d^3{p}}{(2\pi )^3}\rightarrow 2\Sigma _{n=0}^\infty
[2-\delta _{n0}]\int \frac{eB_z}{(2\pi )^2}dp_z,
\]
where the symbol $\delta _{n0}$ denotes the Kronecker delta
\cite{mil14} and thus the spin degeneracy equals 1 for the ground
($n=0$) Landau level and 2 for $n\geq 1$. The redefined density of
states makes the distinctive difference between the magnetic and
non-magnetic cases. The equation (\ref{eq1}) implies that for
$n=0$, $E_{0}=\sqrt{p_{z}^{2}+m_{e}^{2}}$ whereas  for $n\geq 1$
$E_{n}=\sqrt{p_{z}^{2}+m_{e}^{\star 2}}$. These relations indicate
that the quantity $m_{e}^{\star}$ defined as $m_{e}^{\star
2}=m_{e}^{2}+2eB_{z}n$ depends on the presence of the magnetic
field and can be interpreted as an effective electron mass which
is different from electron mass for $n\geq 1$. The number density
of electrons at zero temperature is given by
\begin{eqnarray*}
n_{e}=\Sigma _{n=0}^{n=n_{max}} [2-\delta _{n0}] eB2p_{e}^{F}.
\end{eqnarray*}
The maximum Landau level is calculated from the condition
$(p_{e}^{F})^{2}\geq 0$.
 One can define the critical magnetic density $\rho_{B}$, which
 denotes the limiting density
\begin{equation}
\rho _B=0.802 \ Y_{e}^{-1}\gamma^{3/2} g \  cm^{-3},
\label{eqrhob}
\end{equation}
 for densities lower than
$\rho_{B}$ only the ground Landau level is present.\newline
Thermodynamic properties of a free electron gas in the magnetic
fields at finite temperature have to be investigated. The
temperature affects the electron motion in external magnetic
fields. Finite temperature and decreasing value of the strength of
magnetic field tend to smear out Landau levels.
 The number
density of electrons in the presence of a magnetic field can be
expressed now as
\begin{equation}
n_e=\frac{2m_e^3\gamma }{4\pi ^2}\Sigma _{n=0}^\infty [2-\delta
_{n0}](I_{0,0,+}(z/t,1+2\gamma n)-I_{0,0,-}(z/t,1+2\gamma n))
\end{equation}
where the Fermi integral
\begin{equation}
I_{\lambda ,\eta \pm }(u,\alpha )=\int \frac{(\alpha
+x^2)^{\lambda /2}x^\eta dx}{e^{\left( \sqrt{\alpha +x^2}\mp
u\right) }+1}
\end{equation}
was used \cite{kagan,lai}, $z=\mu _0/m_e$, $t=k_BT_0/m_e$, $u=z/t$
and $\gamma =B_z/B_c$. Similarly to Fermi temperature which is
given by
\begin{equation}
T_F=E_F/k_B=(m_e/k_B)\epsilon _F\qquad (\mathrm{for}%
~~\rho <\rho _B),  \label{eqtf}
\end{equation}
one can define a magnetic temperature
\begin{equation}
T_B={\frac{\Delta E_B}{k_{B}}}={\frac{m_e}{k_{B}}}\left(
\sqrt{1+2n_{\mathrm{max}}\gamma +2\gamma
}-\sqrt{1+2n_{\mathrm{max}}\gamma }\right)  \label{eqtb}
\end{equation}
where $\Delta E_B$ is the energy difference between the
$n_L=n_{\mathrm{max}}$ level and the $n_L=n_{\mathrm{%
max}}+1 $ level. For $\rho =\rho _B$ these temperatures equal
$T_F=T_B$. Knowing the value of magnetic temperature $T_{B}$ one
can describe how the properties of free electron gas changes at
finite temperature and the presence of magnetic field. The
influence of magnetic field is most significant for $\rho \leq
\rho _B$ and $T\leq T_{B}$ when electrons occupy the ground Landau
level. In this case one can deal with the strong quantizing gas
and magnetic field modifies all parameters of the electron gas.
For example, for degenerate, nonrelativistic electrons the
pressure is proportional to $\rho^{3}$ and this form of the
equation of state one can compare with that for the case B=0 for
which $P\sim \rho^{5/3}$. When $\rho > \rho_{B}$ the Fermi
temperature is still greater than magnetic temperature $T_{B}$
electrons are degenerate and there are many Landau levels, now the
level spacing exceeds $k_{B}T$. The properties of the electron gas
are only slightly affected by magnetic field. With increasing
temperature, there is the thermal broadening of Landau levels,
when $T\geq T_{B}$, the free field results are recovered. For
$T\gg T_{B}$ there are many Landau levels and the thermal widths
of the Landau levels are higher than the level spacing. The
magnetic field does not affects the thermodynamic properties of
the gas  \cite{lai}.

The total pressure of the system can be described as the sum of
the pressure coming from electrons and ions plus small corrections
coming from electromagnetic field
\[
P=P_e+P_{ion}+P_{QED}.
\]
The contributions from the same constituents form the energy
density
\[
\varepsilon =\varepsilon _e+\varepsilon _{ion}+\varepsilon _{QED}.
\]
The electron pressure and energy density are defined with the use
of Fermi integral
\begin{equation}
P_e=\frac{2\gamma m_e^4}{4\pi ^2}\Sigma _{n=0}^\infty [2-\delta
_{n0}](I_{-1,2,+}(z/t,1+2\gamma n)+I_{-1,2,-}(z/t,1+2\gamma n))
\end{equation}
\newline
\begin{equation}
\varepsilon _e=\frac{2\gamma m_e^4}{4\pi ^2%
}\Sigma _{n=0}^\infty [2-\delta _{n0}](I_{1,0,+}(z/t,1+2\gamma
n)+I_{1,0,-}(z/t,1+2\gamma n)).
\end{equation}
The ions are very heavy and treated as classical gas. The
influence of magnetic field for ions is very small so we have
neglected this correction. The pressure and energy density
dependence on the chemical potential $\mu _0\,\,$ determines the
form of the equation of state, which is calculated in the flat
Minkowski space-time. The obtained form of the equation of state
is the base for calculating macroscopic properties of the star. In
order to construct the mass-radius relation for given form of the
equation of state the OTV equations have to be solved

\begin{eqnarray}
\frac{dP(r)}{dr}=-\frac{G}{r^2}(\rho (r)+P(r))\frac{(m(r)+4\pi
P(r)r^3)}{(1- \frac{2Gm(r)}{r})}  \label{teq1}
\end{eqnarray}
\begin{eqnarray}
\frac{dm(r)}{dr}=4\pi r^2\rho (r).  \label{teq2}
\end{eqnarray}
However, presence of strong gravitational field of the star causes
the dependence of the temperature and chemical potential on the
gravitational potential. In this paper together with the
calculated form of the equation of state for completely ionised
pure helium plasma the chosen values of central density $\rho_{c}$
changes from $10^{7}$ $kg$ $m^{-3}$ to $10^{14}$ $kg$ $m^{-3}$.
For such parameters the value of magnetic field and temperature
were limited to the following ranges $T(0-10^{9})$ $K$,
$\gamma(0-1)$, respectively. Having solved OTV equations the
pressure $P(r)$, mass $m(r)$ and density $\rho(r)$ were
constructed. To obtain the total radius $R$ of the star the
fulfillment of the condition $P(r)=0$ is necessary. The results
are presented in figures 1-10. Studying the properties of the star
in the framework of the general-relativistic Thomas-Fermi model
one can made the assumption that the temperature and chemical
potential are metric dependent local quantities and the
gravitational potential instead of Poisson's equation satisfies
Einstain's field equations. The metric is static, spherically
symmetric and asymptotically flat
\begin{equation}
g_{\mu \nu }=\left(
\begin{array}{cccc}
e^\nu  & 0 & 0 & 0 \\
0 & -e^\lambda  & 0 & 0 \\
0 & 0 & -r^2 & 0 \\
0 & 0 & 0 & -r^2sin^2\theta
\end{array}
\right) .  \label{tensor metryczny}
\end{equation}
Its coefficients can be determined from Einstain equations and
written as follows
\begin{eqnarray*}
e^{-\lambda }=1-\frac{2Gm(r)}r.  \label{leq}
\end{eqnarray*}
The energy-momentum conservation
\[
T_{;\nu }^{\mu \nu }=0
\]
which for spherically symmetric metric (\ref{tensor metryczny})
can be written as
\begin{eqnarray}
\frac{d\nu }{dr}=-\frac 2{P+\rho }\frac{dP}{dr}  \label{teq}
\end{eqnarray}
together with the Gibbs-Duhem relation and with the assumption
that the heat flow and diffusion vanishes \cite{isr} give the
condition
\begin{eqnarray}
\frac \mu T=\mathrm{const}.  \label{eq19}
\end{eqnarray}
This implies that the temperature and chemical potential be come
local metric functions
\begin{eqnarray}
T(r) &=&e^{-\nu (r)/2}T_0  \label{tol1} \\
\mu (r) &=&e^{-\nu (r)/2}\mu _0  \label{tol2}
\end{eqnarray}
where $T_0$ and $\mu _0$ are constants equal to the temperature
and chemical potential at infinity. The temperature $T_0$ may be
chosen arbitrarily as the temperature of the heat-bath. First
equation in (\ref{tol1}) is the well known Tolman condition for
thermal equilibrium in a gravitational field \cite {tol}. For
given values $\mu_{0}$ and $T_{0}$ one can obtain self-consistency
equations defining general-relativistic Thomas-Fermi equation.
Now, the Fermi-Dirac distributions are defined as \cite{bilic1}
\begin{eqnarray}
& \frac 1{e^{\left( \sqrt{1+x^2+2\gamma n}-z_0\right) \slash %
t_0}+1}\rightarrow \frac 1{e^{\left( e^{\nu (r)/2}\sqrt{1+x^2+2\gamma n}%
-z_0\right) \slash t_0}+1} \nonumber \\
 & =\frac 1{e^{\left( \sqrt{1+x^2+2\gamma n}%
-z(r)\right) \slash t(r)}+1}. & \nonumber
\end{eqnarray}
In the result instead of the OTV equilibrium equations together
with the equation of state which was calculated in the flat
space-time one can derived three self-consistency equations
(\ref{teq1},\ref {teq2},\ref{teq}) together with local form of the
equation of state being now the function of $r$. This is of
particular importance in the case of the strong gravitational
fields.

\section{Discussion}

In order to construct the mass-radius relation for white dwarfs
the proper form of the equation of state have to be enumerated.
For our needs we have chosen the pure helium plasma as the main
constituent of the white dwarf interior. All calculations were
performed at finite temperature and different from zero magnetic
field and compared with those of zero temperature and without
magnetic field. In Figure 1 the applied form of the equation of
state for magnetic and non-magnetic white dwarfs were presented.
The zero temperature case is compared with the one obtained for
temperature $6\times 10^{8}$ $K$. For temperature equals zero
there are clearly visible Landau levels which are smeared out when
the temperature is different from zero and the strength of
magnetic field is decreased. The density profile for helium
magnetic and non-magnetic white dwarfs for different values of
temperatures and central density $\rho_{c}$ equals $10^{10}$ $kg$
$m^{-3}$ was shown in Figure 2. As in the case of the EOS one can
distinguish Landau levels for zero temperature case. The Fermi
temperature for non-magnetic white dwarfs with the fixed value of
the central density $\rho_{c}=10^{10}$ $kg$ $m^{-3} $ is
calculated and equals $T_{F}=1.023\times10^{10}$ $K$. The
estimated value of the magnetic field temperature for white dwarfs
with the strength of magnetic field $\gamma=1$ $(\rho_{c}=10^{10}
\  kg \  m^{-3} )$ is $T_{B}=1.27\times10^{9}$ $K$. Theoretical
model of helium white dwarf with different electron masses which
correspond to the ground Landau level $(n=0)$ and the one for
$n=1$ is presented in Figure 2. The change of mass can be
interpreted as the appearance two shells with distinctively
different value of densities. For the presented form of the EOS
the mass-radius relations were constructed. These relations are
presented in Figure 3. Of special interest is the case when the
temperature equals $10^8$ $K$ because the appearance of unstable
areas. The same is visible on the relation mass-central density
(Figure 4). There are marked points $A$, $B$, $C$ on mass-radius
and mass-central density relations. These points represents the
stable configurations ($A$ and $C$) whereas point $B$ refers to
the unstable one. Figures 5 and 6 present sequences on the
mass-radius relations for fixed value of temperature. In Figure 5
the chosen value of the temperature equals $T=6\times 10^{8}$ $K$
whereas in Figure 6 the temperature equals $T=3\times10^{8}$ $K$ .
In
both figures the strength of magnetic field changes from $\gamma =0$ to $%
\gamma =1$. For comparison the zero temperature mass-radius
relation obtained was presented. For increasing value of the
magnetic field white
dwarfs increase their mass. However, the asymptotic value of $Chadrasekhar$ $%
mass$ is not exceeded. In the range of low densities together with
increasing value of the temperature the radii of these objects
decrease. The mass-central density relations for helium white
dwarfs are presented in Figures 7 and 8. Thick full curves stable
branches, whereas thin full curves depict the unstable branches
(Figure 8). The minimum and maximum of helium white dwarf star
masses are represented by the extrema of $M$. These are point
where the stable and dynamically unstable branches merge. Stable
branches are those with $\frac{dM}{d\rho _c}>0$. The existence of
unstable branches enables different evolution tracks. The
temperature profile is presented in the Figure 9 for the fixed
value of the central density $\rho _{c}$. The temperature is
changed slightly with the radius in agreement with the value of
gravitational potential which is much smaller in comparison with
neutron stars. White dwarfs with known values of masses and radii
which collected in Table 1 are marked on the obtained mass-radius
relations in Figure 10. In Figure 10 the theoretical point $m$ is
marked. It represents the model of a star formed with
one-dimensional electron gas. The changes of electron mass depend
on the value of the Landau level $n$. The constructed
configuration has distinctive shells with different values of the
electron mass which is directly connected with the population of
Landau levels. For $n=0$ (only the ground Landau level is
populated) the electron mass is unchanged whereas for $n=1$, one
can deal with the modified by the magnetic field electron mass.

\newpage
\section{Figure captions.}

Figure 1
\newline 
The equation of state for different temperature
and strength of magnetic field cases. 
\newline 
Figure 2
\newline
The density profile for helium magnetic white dwarf and for
non-magnetic white dwarf for different temperatures and
$\rho_{c}=10^{10}$ $kg/ m^{-3}$.
\newline 
Figure 3
\newline
The mass-radius relation for white dwarf with different temperature cases.%
\newline
Figure 4
\newline 
The mass $M$ dependence on the central density
$\rho_{c}$ for different temperature values.
\newline 
Figure 5
\newline 
The mass-radius diagram for helium white dwarfs for
temperature $T=6$ $10^{8} $ $K$ and the strength of magnetic field
$\gamma=0.1-1$. The solid line denotes the mass-radius relation
for $T=0K$.
\newline 
Figure 6
\newline 
The mass-radius diagram for
helium white dwarfs for temperature $T=3$ $10^{8} $ $K$ and the
strength of magnetic field $\gamma=0.1-1$.
\newline 
Figure 7
\newline
The mass $M$ dependence on the central density $\rho_{c}$ for temperatures $%
T=0K$ and $T=6$ $10^{8}$ $K$.
\newline Figure 8
\newline
The mass $M$ dependence on the central density $\rho_{c}$ for temperatures $%
T=0K$ and $T=3$ $10^{8}$ $K$.
\newline 
Figure 9
\newline 
The temperature profile for different central density. 
\newpage 
Figure 10
\newline 
The mass-radius relation for white dwarf with different
values of temperature with marked observational white dwarfs. The
point $m$ denotes the theoretical model of the star with Landau
levels $n_{L}=0$, $n_{L}=1$ which are clearly visible in the
Figure 2. The mass and radius for this configuration is calculated
$R=0.92$ $(100R/R_{\odot})$, $M=1.18$ $M_{\odot}$.
\newpage

\begin{figure}
{}
\par
\centering \resizebox*{12cm}{!}{\includegraphics{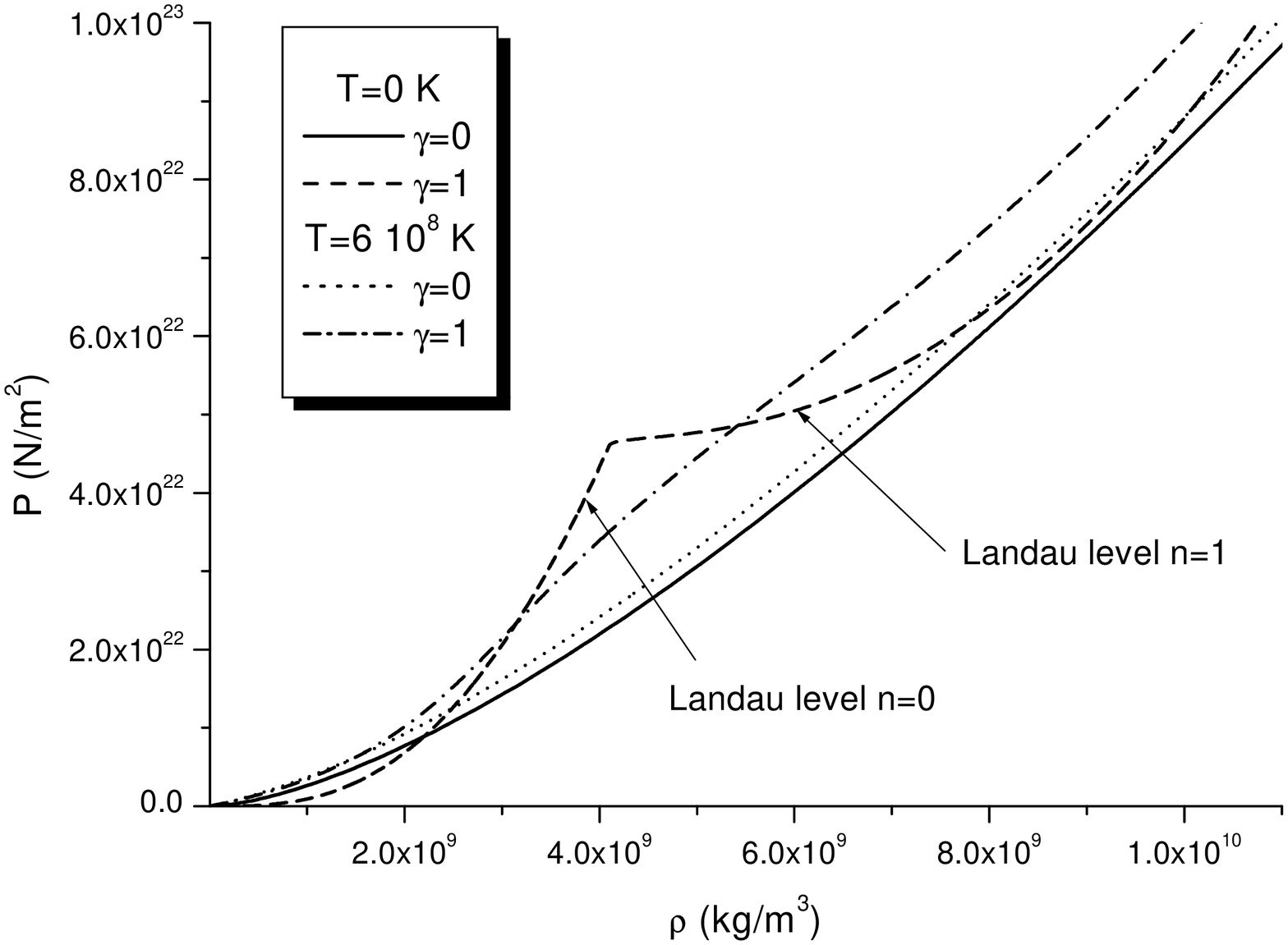}}
\par
\caption{}
\end{figure}
\samepage
\begin{figure}
{}
\par
\centering \resizebox*{12cm}{!}{\includegraphics{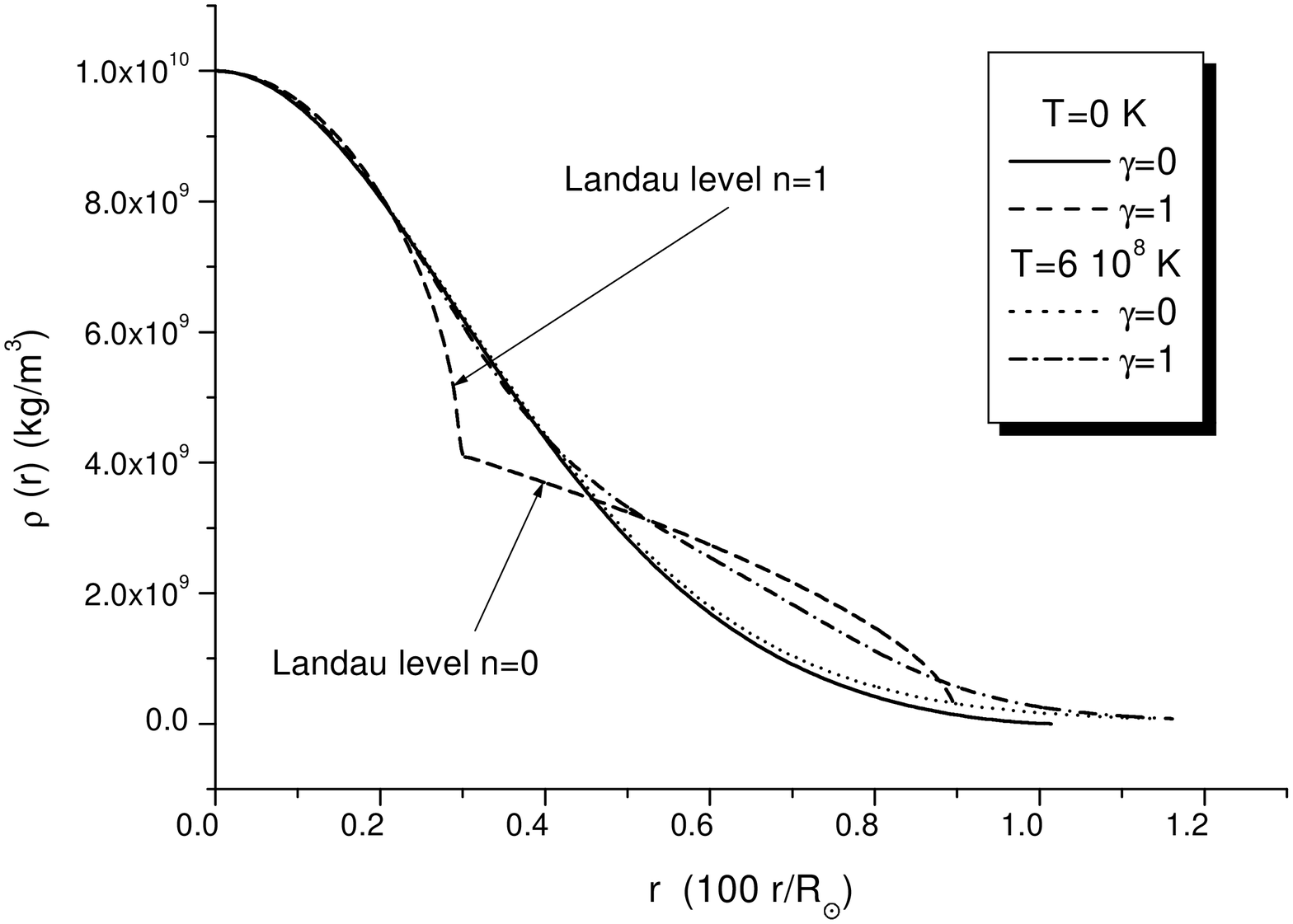}}
\par
\caption{}
\end{figure}
\pagebreak
\begin{figure}
{}
\par
\centering \resizebox*{12cm}{!}{\includegraphics{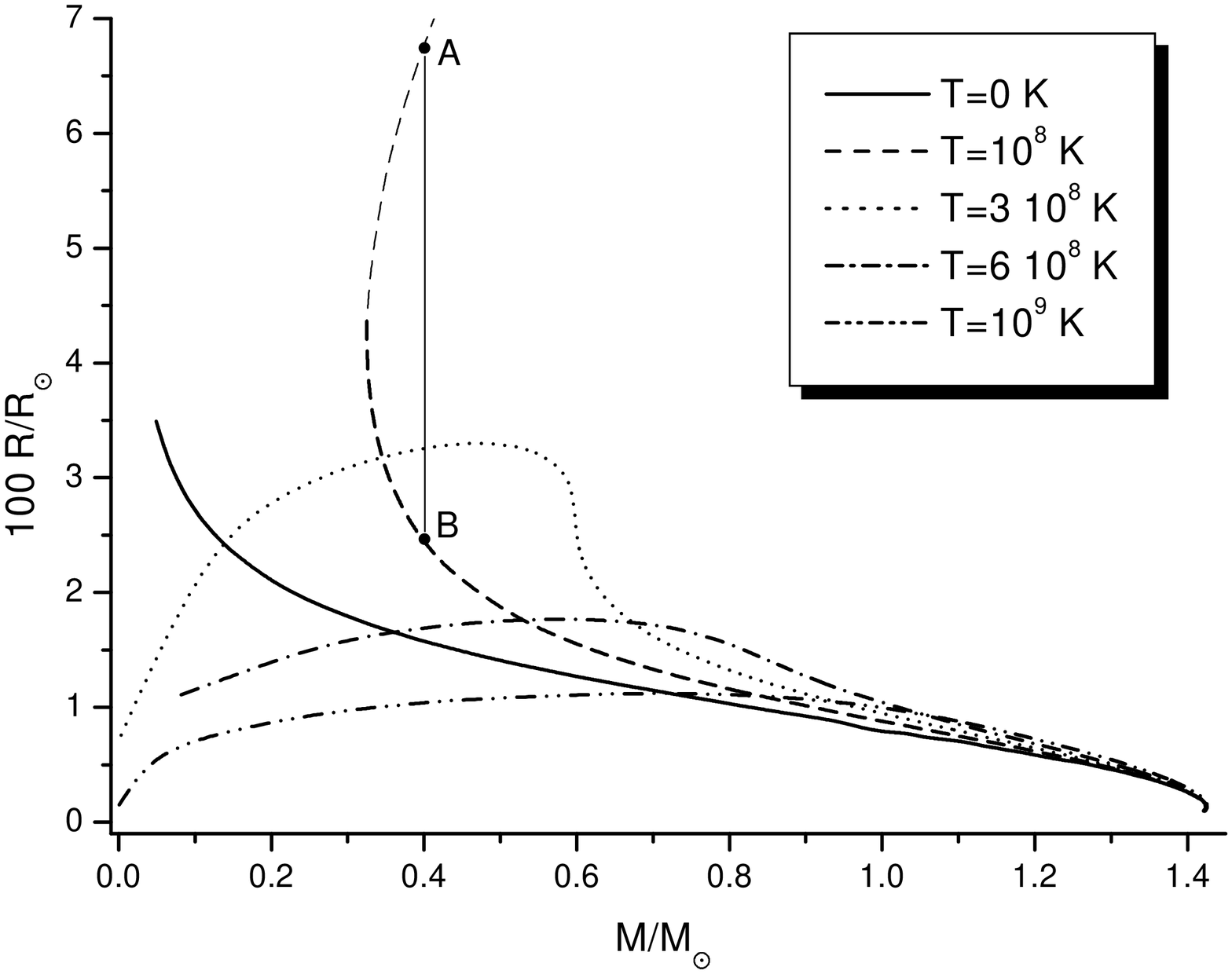}}
\par
\caption{} \label{rys 3}
\end{figure}
\samepage
\begin{figure}
{}
\par
\centering \resizebox*{12cm}{!}{\includegraphics{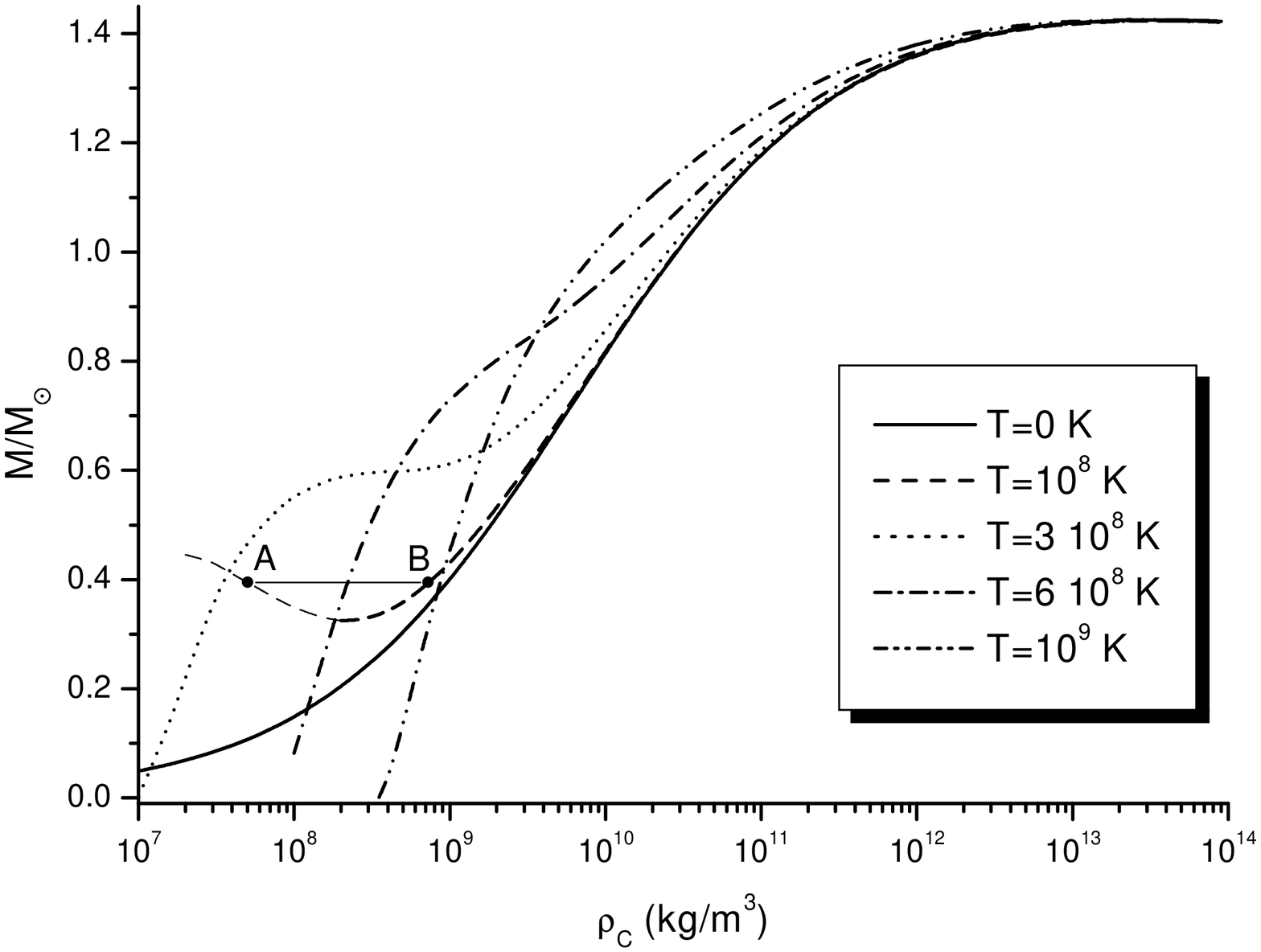}}
\par
\caption{} \label{rys 4}
\end{figure}
\pagebreak
\begin{figure}
{}
\par
\centering
\resizebox*{12cm}{!}{\includegraphics{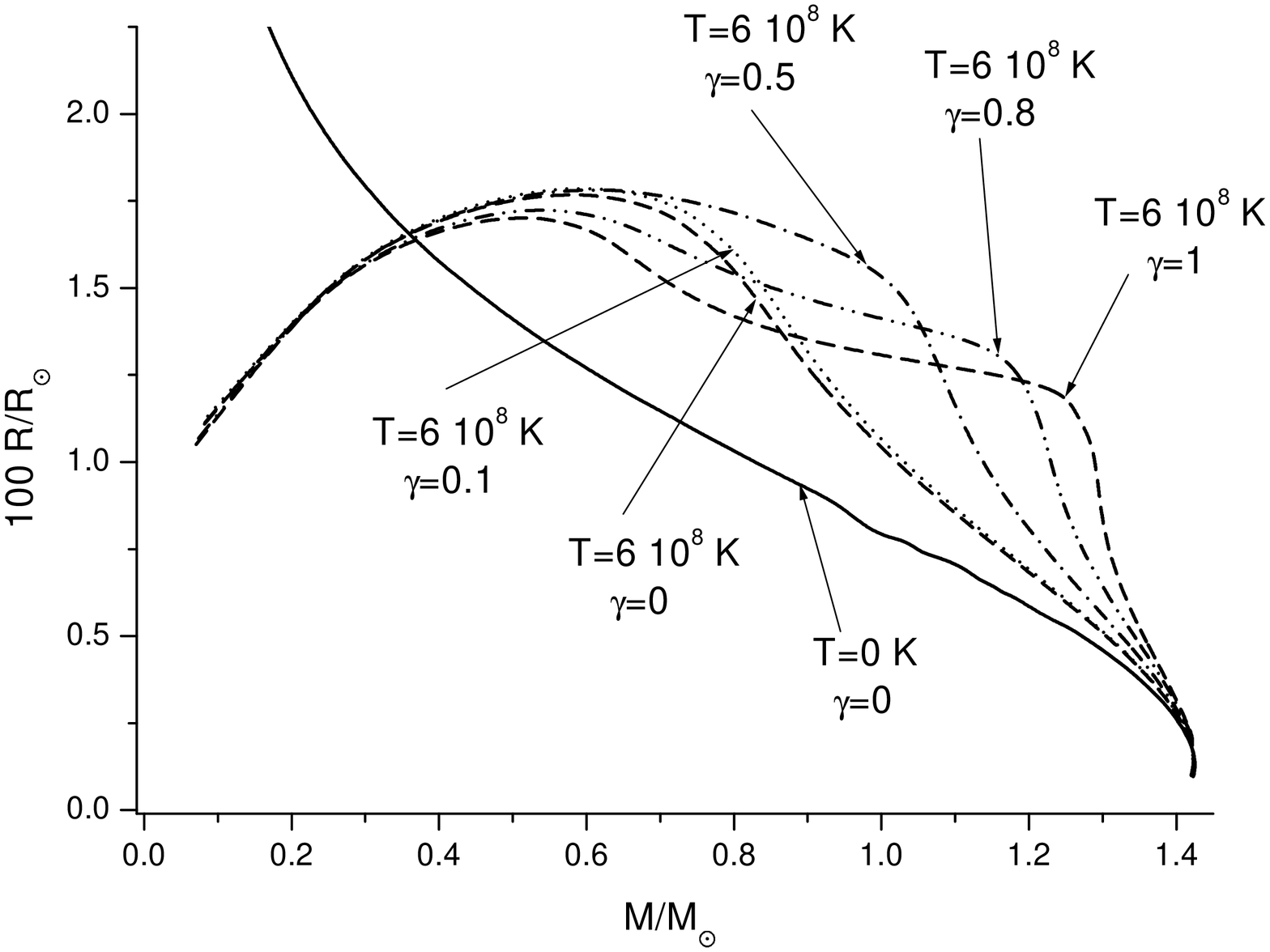}}
\par
\caption{} \label{rys 5}
\end{figure}
\samepage
\begin{figure}
{}
\par
\centering
\resizebox*{12cm}{!}{\includegraphics{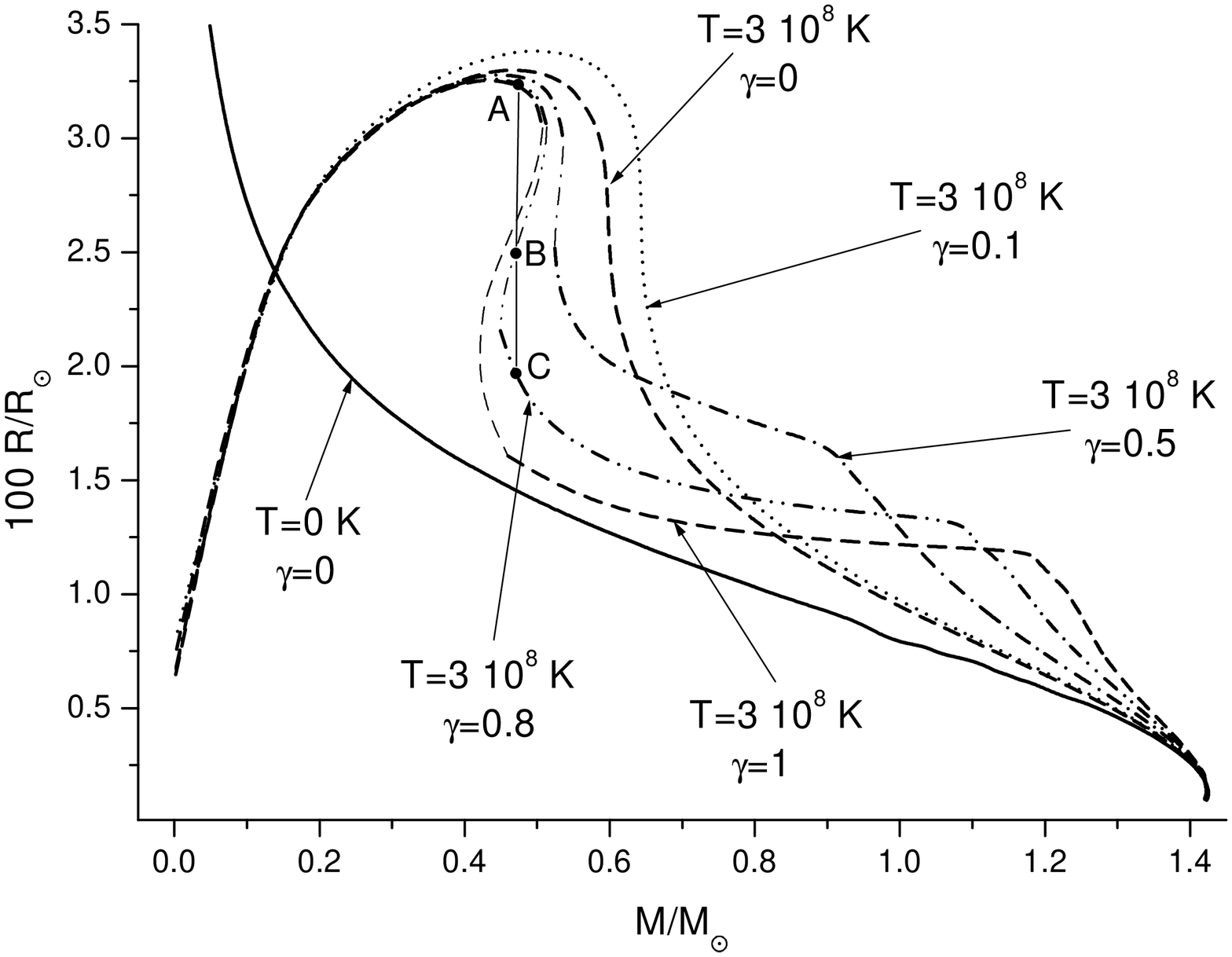}}
\par
\caption{} \label{rys 6}
\end{figure}
\pagebreak
\begin{figure}
{}
\par
\centering
\resizebox*{12cm}{!}{\includegraphics{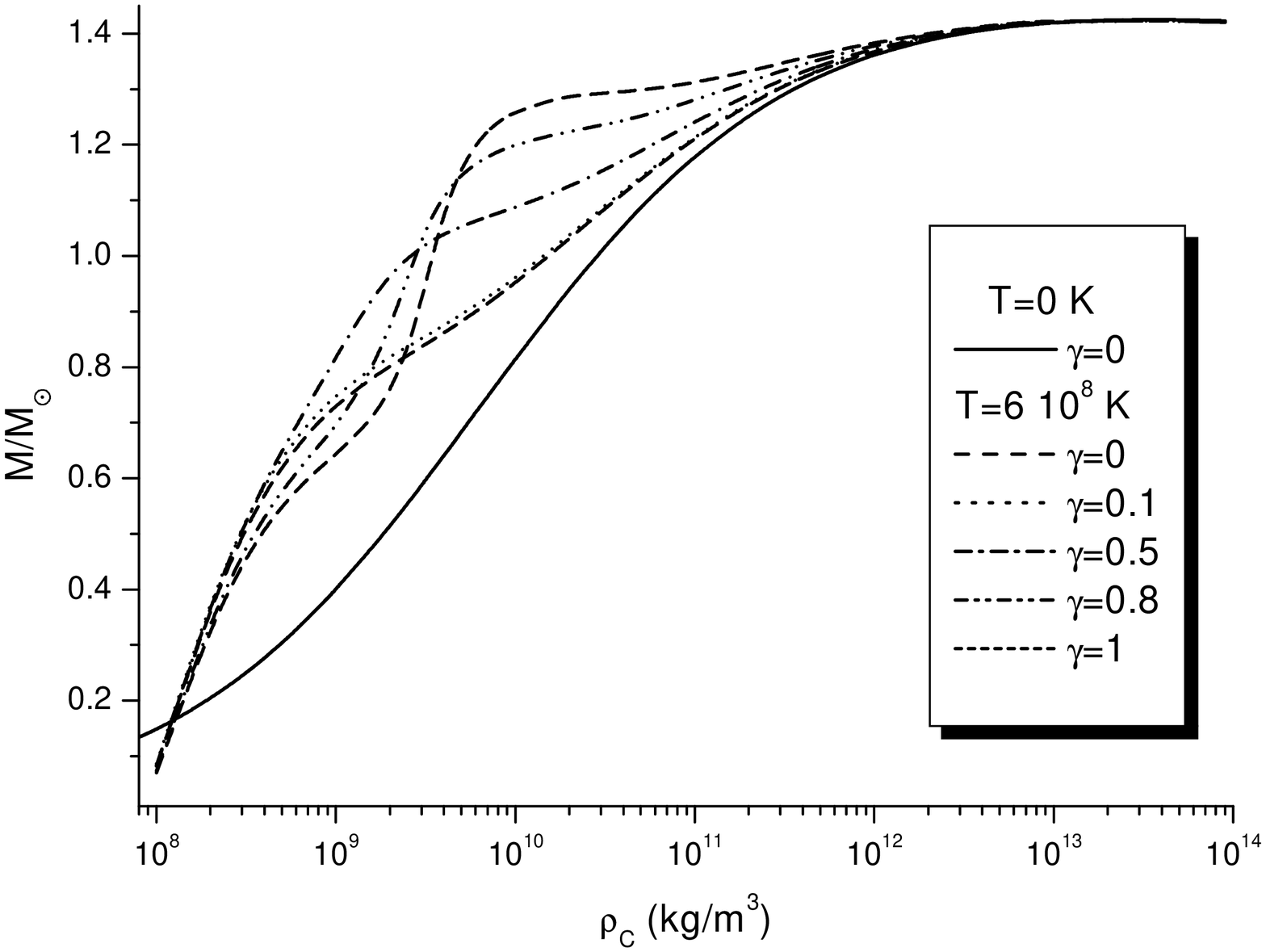}}
\par
\caption{} \label{rys 7}
\end{figure}
\samepage
\begin{figure}
{}
\par
\centering
\resizebox*{12cm}{!}{\includegraphics{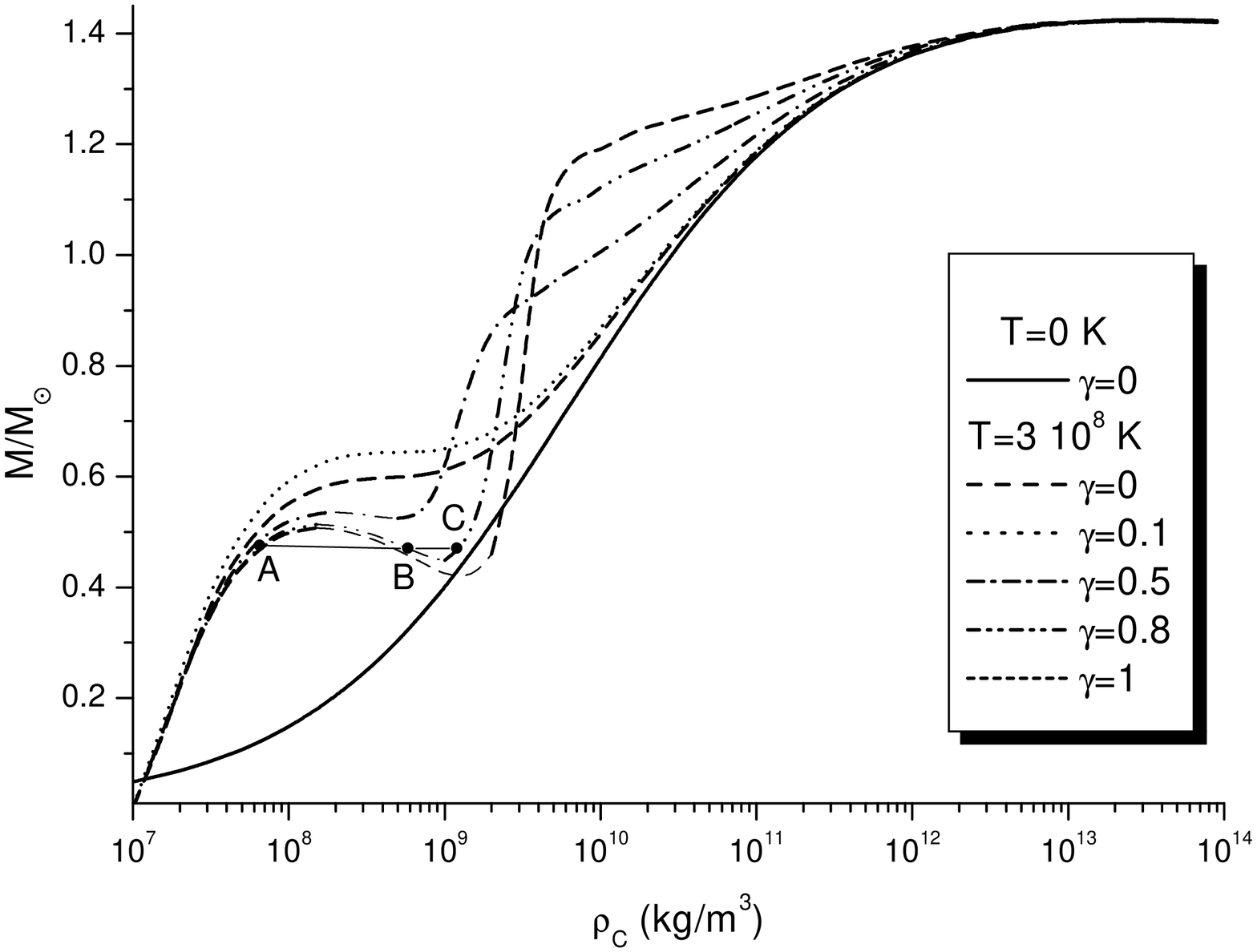}}
\par
\caption{} \label{rys 8}
\end{figure}
\pagebreak
\begin{figure}
{}
\par
\centering \resizebox*{12cm}{!}{\includegraphics{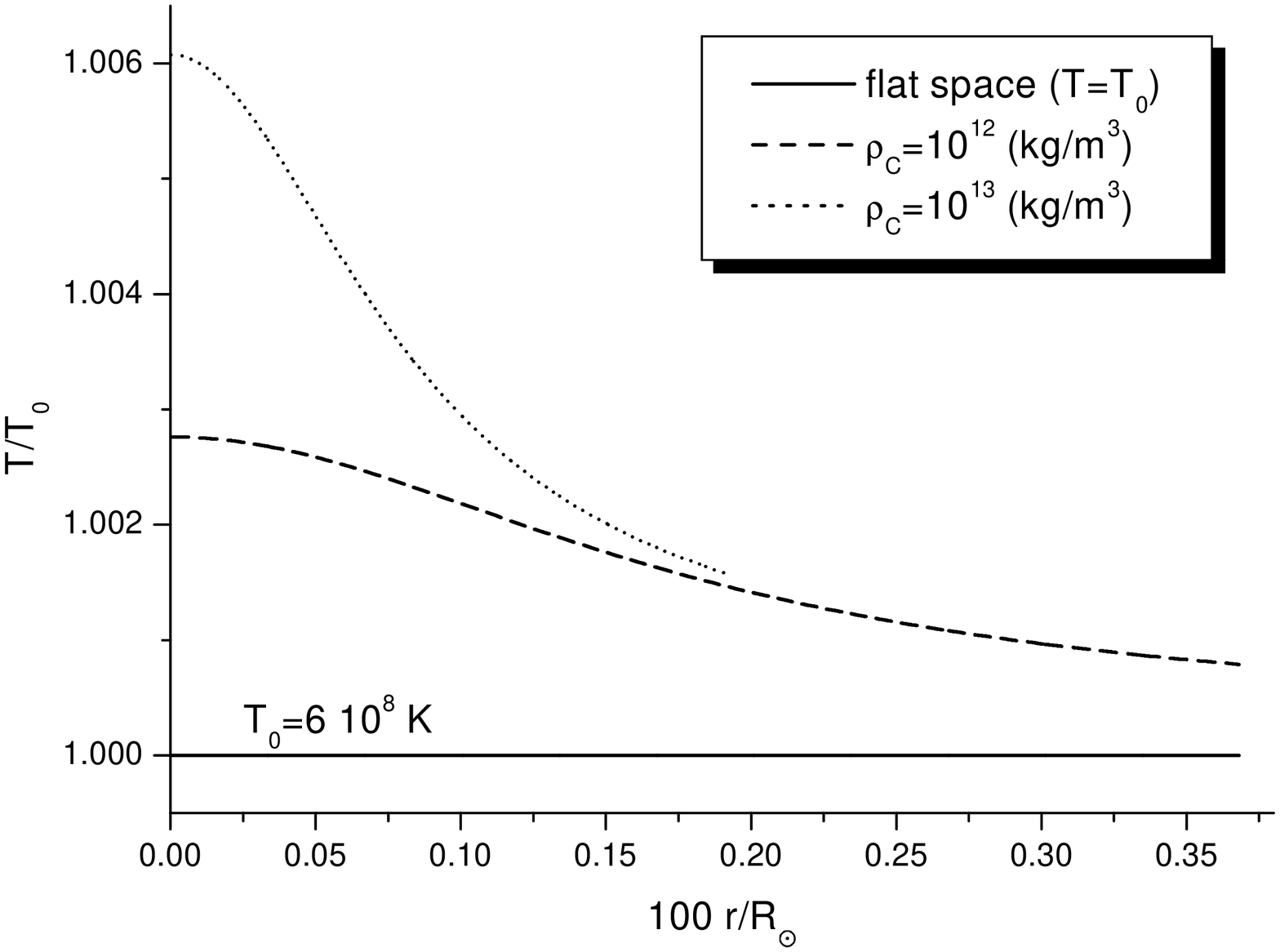}}
\par
\caption{} \label{rys 9}
\end{figure}
\samepage
\begin{figure}
{}
\par
\centering
\resizebox*{12cm}{!}{\includegraphics{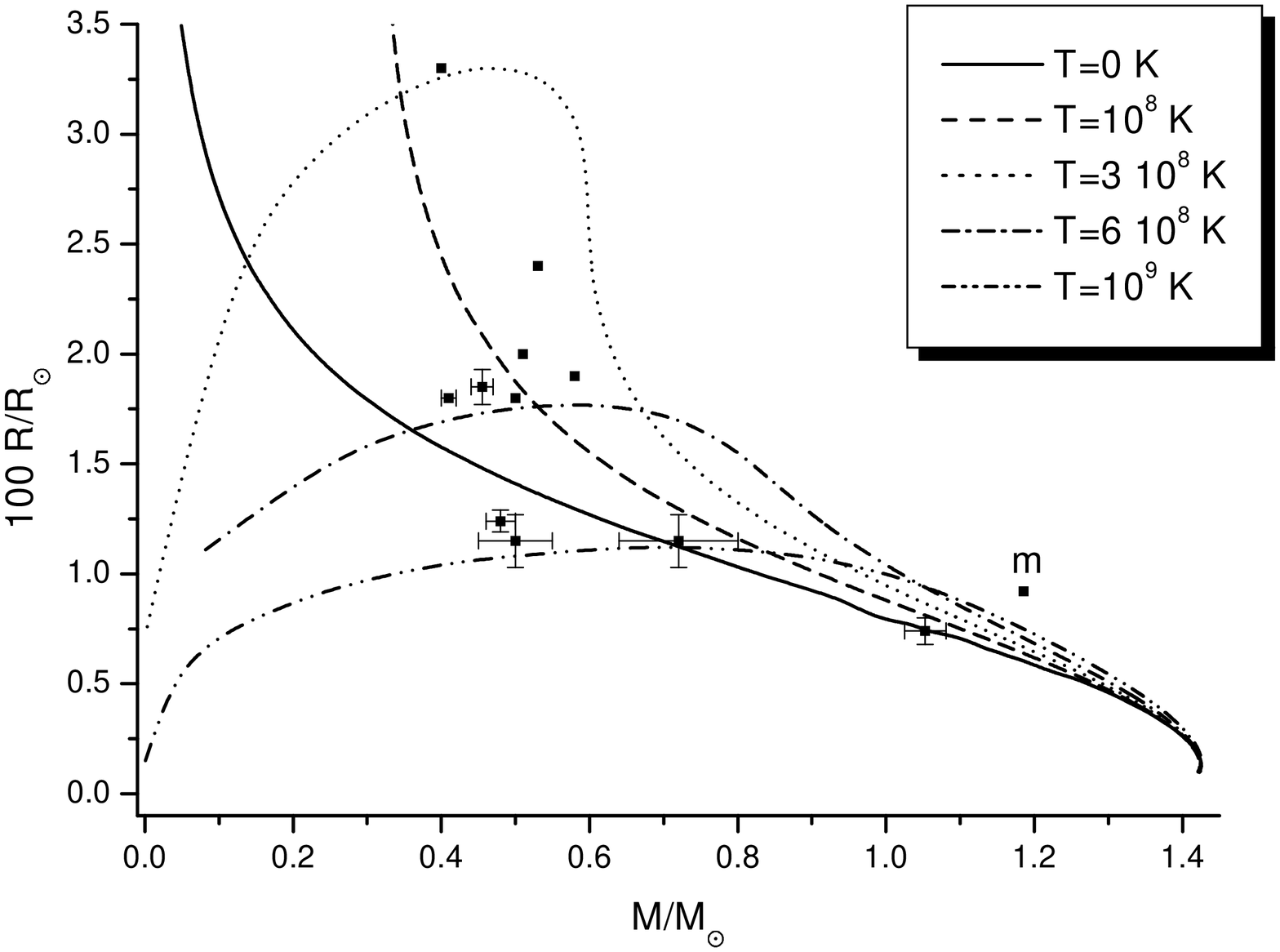}}
\par
\caption{} \label{rys 10}
\end{figure}

\newpage

\begin{table}[tbp]
{\centering
\begin{tabular}{|c|c|c|c|}
\hline $Object$ & $M\, [\, M_{\odot }\, ] $ & $R\, [\, \frac{100
R}{R_{\odot}}\, ] $ & $Ref.$
\\ \hline\hline
$Syriusz $B%
$$
& $ 1.053\pm0.028 $ & $ 0.74\pm 0.06 $  & $ \cite{kot5}$\\
\hline
$$
40%
$$
Eri%
$$
B%
$$
& $ 0.48\pm 0.02 $ & $ 1.24\pm0.05 $ & $ \cite{kot5}$ \\
\hline
$$
Stein%
$$
2051%
$$
& $0.50\pm 0.05 $ $or$ $0.72\pm 0.08 $ & $1.15\pm 0.12 $ & $\cite{kot5}$ \\
\hline
$$
DA%
$$
& $0.45\pm0.015 $ & $1.85\pm 0.08 $ & $\cite{ben}$  \\
\hline
$$
GD448%
$$
& $0.41\pm0.01 $ & $1.8 $ & $\cite{max}$ \\ \hline
$$
GD191-B2B%
$$
& $0.51 $ & $2.0 $ & $\cite{ban}$ \\ \hline
$$
Feige%
$$
& $0.50 $ & $1.8 $ & $\cite{ban}$ \\ \hline
$$
WD2218+706%
$$
& $0.40 $ & $3.3 $  & $\cite{ban}$\\ \hline
$$
REJ0558+165%
$$
& $0.58 $ & $1.9$  & $\cite{ban}$\\ \hline
$$
REJ1738+665%
$$
& $0.53 $ & $2.4 $  & $\cite{ban}$\\ \hline
\end{tabular}
}
\par
\caption{The masses and radii for some typical white dwarfs.}
\label{tab: 1}
\end{table}

\end{document}